\documentclass[prd,superscriptaddress, showkeys, preprintnumbers,nofootinbib]{revtex4-2}
\usepackage{graphicx}
\usepackage{amsmath}
\usepackage{amssymb}
\usepackage{mathrsfs}
\usepackage{tikz}

\usepackage[caption = false]{subfig}
\usepackage[normalem]{ulem}
\usepackage{xcolor}
\definecolor{lcolor}{rgb}{0.5,0,0}
\definecolor{citcolor}{rgb}{0,0.3,0.0}

\usepackage[colorlinks=true,breaklinks=true]{hyperref}
\hypersetup{allcolors=[rgb]{0.0 0.0 0.75},linkcolor=[rgb]{0.75 0.05 0.05}}
\usepackage{mciteplus}

\begin{document}
\title{Looking for Condensed Gluons: A Cross-Scale Journey from the Deep Structure of Protons to High-Energy Cosmic Rays—A Mini-Review}
%\pacs{24.85.+p, 13.60.-r}

\preprint{}
\author{Wei Zhu}
\affiliation{Center for Fundamental Physics, School of Artificial Intelligence, Hubei Polytechnic University, Huangshi 435003, China;}
\affiliation{Department of Physics, East China Normal University, Shanghai 200241, China;}
%\email{yeyin.zhao@gmail.com}
\author{Yu-Chen Tang}
\affiliation{College of Science, Westlake University, Hangzhou 310030, China;}

\author{Ye-Yin Zhao}
\affiliation{ School of Physics and Electronic Engineering,  Sichuan University of Science and Engineering, Zigong 643000, China}

\author{Bo Yang}
\affiliation{Center for Fundamental Physics, School of Artificial Intelligence, Hubei Polytechnic University, Huangshi 435003, China;}
\author{Yu-Chen Xiong}
\affiliation{Center for Fundamental Physics, School of Artificial Intelligence, Hubei Polytechnic University, Huangshi 435003, China;}
\date{\today}

\begin{abstract}
    Quark--gluon dynamics within protons and high-energy radiation phenomena in the universe are typically regarded as two entirely distinct fields. This paper aims to demonstrate that gluon condensation (GC) may serve as a direct bridge between these two fields. We review three key aspects of GC research: first, the Zhu--Shen--Ruan (ZSR) equation, as a nonlinear evolution equation based on structural symmetry, exhibits self-consistent connections with the DGLAP, BFKL and GLR-MQ-ZRS equations, providing a theoretical foundation for the generation of GC; 
second, the chaotic solutions and the shadowing--antishadowing synergy inherent in this equation can drive gluons to aggregate near the critical momentum, thereby forming a novel type of high-density, strongly interacting matter; third, these changes in microstructure manifest themselves as a broken-power-law feature in high-energy cosmic gamma-ray spectra, thereby offering new insights into the hadronic scenarios underlying certain astrophysical sources. Consequently, GC not only concerns the novel behaviour of quantum chromodynamics under extreme conditions but may also serve as a vital window for probing the deep structure of protons using cosmic-ray signals. With the advancement of higher-precision gamma-ray observations, hadron collision experiments and related theoretical research, the physical picture of GC and its observational criteria are expected to undergo more rigorous testing. Should this picture be confirmed, certain features in the high-energy gamma-ray spectrum (previously attributed solely to empirical fitting or lepton models) will need to be re-examined within the deeper context of hadronic dynamics; simultaneously, GC may also provide a new entry point for research into pion condensation in nuclear physics and even condensed matter physics. Consequently, the significance of the search for GC extends beyond the model itself, reaching into multiple fields of natural science.
\end{abstract}


\maketitle

\section{Introduction}
\begin{figure}[htbp]
  %\centering
  \includegraphics[width=0.7\textwidth]{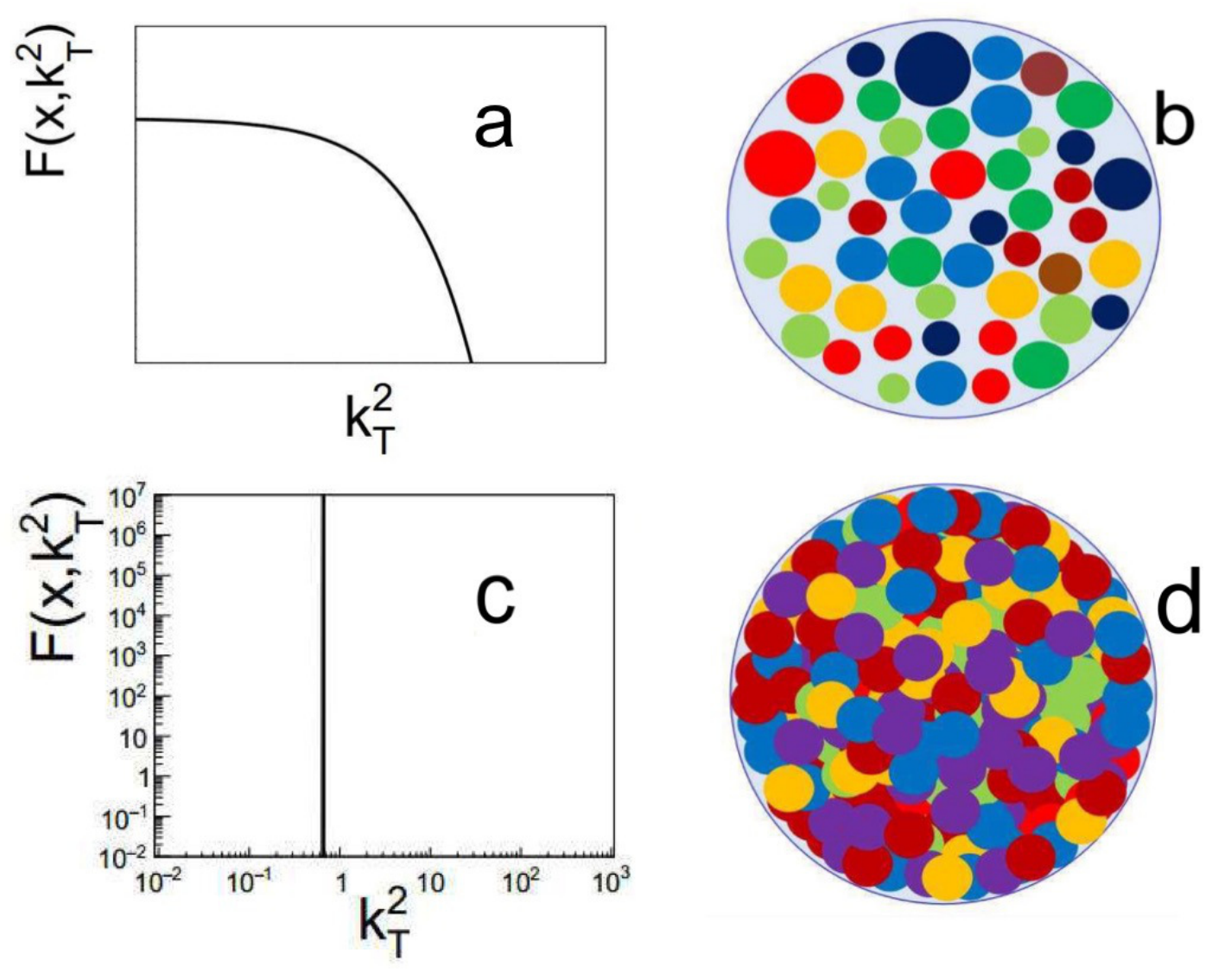}
  \caption{Schematic diagrams of CGC (\textbf{a},\textbf{b}) and GC (\textbf{c},\textbf{d}). {\color{black}{The GC threshold $k_c^2\simeq$ CGC scale $Q^2_s$, but $x_c\ll x_s$.}}}\label{fig:1}   	
\end{figure}

Human understanding of nature now spans an unprecedentedly vast range, extending from the quark--gluon distribution
within protons to cosmic ray signals originating from various galaxies. The search for connections between phenomena occurring at these two extreme scales presents a challenging task that bridges the fields of particle physics and astrophysics~\citep{Torres:2004hk, Muller:2025qof}. %[1,2].
The Quantum Chromodynamics (QCD) evolution equations are an indispensable tool for studying quark--gluon distribution functions~\citep{Sterman1993}. The existence of gluons in the proton was first established by two classic evolution equations: the Dokshitzer--Gribov--Lipatov--Altarelli--Parisi (DGLAP)~\citep{Gribov:1972ri, Dokshitzer:1977sg, Altarelli:1977zs} and Balitsky--Fadin--Kuraev--Lipatov (BFKL)~\citep{Lipatov:1976zz, Fadin:1975cb, Kuraev:1976ge, Kuraev:1977fs, Balitsky:1978ic} equations. Further insights into gluons within the proton are predicted by non-linear modifications of these two equations, namely the Gribov--Levin--Ryskin--Mueller--Qiu (GLR-MQ)~\citep{Gribov:1983ivg, Mueller:1985wy} and Balitsky--Kovchegov (BK)~\citep{Balitsky:1995ub, Kovchegov:1999yj, Kovchegov:1999ua} equations. These emphasise that gluon splitting leads to an increase in gluon density, whereby gluons suppress their own unreasonably rapid growth through aggregation. In particular, the Jalilian-Marian--Iancu--McLerran--Weigert--Leonidov--Kovner (JIMWLK) equations~\citep{Jalilian-Marian:1996mkd, Jalilian-Marian:1997qno, Jalilian-Marian:1997jhx, Iancu:2001md, Weigert:2000gi}, which are a generalisation of the BK equation to the case of multiple gluon correlations, predict a saturation limit for the gluon distribution---the Color Glass Condensate (CGC) (Fig.~\ref{fig:1}a,b)~\citep{McLerran:1993ni, McLerran:1993ka, McLerran:1994vd}---thereby sparking a surge of interest in the search for CGC in heavy-ion colliders~\citep{Garcia-Montero:2025hys, Jalilian-Marian:2005ccm, Gelis:2010nm}. However, CGC has yet to break free from the confines of the accelerator laboratory and venture into the distant depths of the universe, as it remains a relatively weak effect that cannot be detected through astronomical observations.

{This paper sheds light on a peculiar aspect of gluon behaviour within the \mbox{proton---gluon}} condensation (GC), which may provide a unique perspective on the aforementioned research. All QCD evolution equations describe changes in the parton distribution function under the fundamental interactions of QCD (splitting and recombination). Evolutionary kernels (evolutionary dynamics) can be linear or non-linear. Each type exhibits both positive and negative components, which is a consequence of momentum conservation. The contributions from the positive and negative non-linear terms are referred to as antishadowing (positive feedback) and shadowing (negative feedback), respectively. The shadowing effect weakens the particle distribution, whilst the partons whose intensity has been weakened in turn reduce the shadowing effect; this process continues until equilibrium and saturation are ultimately reached. The JIMWLK equation describes precisely this process. Unlike shadowing, antishadowing is a positive feedback mechanism; the stronger it is, the stronger the antishadowing  becomes. Consequently, even after a very short evolution path, the number of partons tends to infinity. In practical examples, the GLR equation employs the AGK cutting rule~\citep{Abramovsky:1973fm} to generate a negative non-linear term, resulting in a net shadowing  effect.  In the BK equation, the non-linear term is generated by the dipole splitting mechanism, which also produces a net shadowing effect. The shadowing  effects of both equations are not strong enough. (Note that the flat distribution observed in the impact parameter space does not indicate saturation!) It is only after the JIMWLK equation sums over multipole amplitudes that the strong shadowing effect produces the saturated distribution---CGC.

It is well known that perturbation theory in quantum field theory requires the summation of contributions from all correlated processes of the same order; this not only eliminates infrared divergences but also preserves momentum conservation. Work~\citep{Zhu:1998hg} utilised the Time-Ordered Perturbative Theory (TOPT) cutting rule to improve the GLR-MQ and BK equations, thereby preserving momentum conservation whilst allowing for the coexistence of shadowing and antishadowing. The resulting equations are denoted as the Zhu--Ruan--Shen (ZRS)~\citep{Zhu:1999ht, Zhu:2004xj} and Zhu--Shen--Ruan (ZSR)~\citep{Zhu:2016qif} equations, respectively.

The ZRS equation introduces an antishadowing correction to the GLR equation; however, as shadowing continues to dominate the evolution, the significance of antishadowing has not yet received sufficient attention~\citep{Wang:2023zop, Lalung:2019tdp, Lalung:2019qvn}. It was not until the ZSR equation, where the Lipatov singularity appears in the nonlinear term, that chaos was induced; the violent oscillations resulting from this chaos, in turn, caused the distribution function at the singularity to exhibit strong antishadowing. In concert with the strong shadowing that occurs simultaneously, the saturated gluons are further compressed into a novel class of condensate---GC (Fig.~\ref{fig:1}c,d). Owing to the highly distinctive behaviour of the condensed gluons, we can use cosmic-ray signals at energies far exceeding those of existing accelerators to peer into the deep structure of the proton; conversely, this pathway also opens a new window for us to understand the origin of cosmic-rays and the evolution of the universe from the deepest levels of matter.

To help readers from different fields understand GC, we aim to provide a straightforward overview of how GC achieves the aforementioned objectives. All QCD evolution equations are approximate descriptions of parton distributions within specific kinematic regions. There ought to be an intrinsic connection between these equations, and they should be consistent in both their physical picture and mathematical structure. Indeed, in Section~\ref{sec2}, when we bring together and compare the four QCD evolution equations describing the gluon distribution function in the small-x region, we find that they exhibit a specific reciprocity.  Three of these equations, DGLAP, BFKL and GLR-MQ-ZRS equations, are the well-known and widely used. In Section~\ref{sec3}, we show that it is precisely the fourth ZSR equation, which is coupled to these three, that can further evolve gluons in the CGC state into GC, thereby unleashing a storm of butterfly effects within the proton.

In Section~\ref{sec4}, we describe our search for GC. Since the GC energy threshold may exceed the LHC's energy range, we look for traces of GC in cosmic rays. The method is straightforward.
We incorporate the GC distribution predicted by the QCD evolution equations into the hadronic scenario of the cosmic gamma-ray spectrum commonly used in astronomy. Through simple elementary calculations, we obtain an analytical solution for the gamma-ray spectrum exhibiting GC characteristics. Applying this to re-examine the observed cosmic gamma-ray spectrum, we have discovered that GC manifests itself extensively in astrophysical phenomena~\citep{Zhu:2017bvp}. It is particularly worth emphasising that, within the chain linking proton structure to cosmic-rays, a crucial link is the assumption that a vast number of condensed gluons leads to saturation of the secondary pion production. This section devotes considerable space to analysing the validity of this assumption.
The final Section offers an outlook on GC. We will demonstrate the potential applications of GC across a wide range of disciplines.

\section{Evolution of the QCD Evolution Equations\label{sec2}}

Before proceeding, we emphasise that the purpose of the present mini-review is not to reproduce the full technical derivations developed in our earlier systematic studies, but to extract three key elements of the GC framework and present them in a more accessible form for readers from different fields. The detailed derivation of the ZSR equation, the numerical analysis of its chaotic solutions, and the formulation of the GC-induced gamma-ray spectrum have all been discussed extensively elsewhere. Here we retain only the minimum theoretical structure needed to make the logical connection among these ingredients transparent across the scales from proton structure to high-energy astrophysical observations. We therefore do not intend this abbreviated presentation to substitute for the original technical papers; rather, it is meant as a synthetic and pedagogical overview of a broader line of work.

The ZSR equation can be derived within the TOPT framework using standard perturbation theory methods. However,
this section derives it using a more intuitive physical picture: the proton is composed of quarks and gluons (partons). The QCD factorisation theory decomposes high-energy proton processes into perturbative and non-perturbative parts. The application of the renormalisation group and its parton picture~\citep{Narison:2002woh} yields a relationship that is universally present between the perturbative QCD evolution kernels and the non-perturbative distribution functions, namely the QCD evolution equations. In different energy regions, there are different approximations for the evolution kernels and the objects of evolution (distribution functions); their specific forms are model-dependent. Identifying the intrinsic connections between different evolution kernels is crucial for exploring new QCD evolution equations. We refer to this as evolution of the QCD evolution~equations.

\begin{figure}[htbp]

    \includegraphics[width=.7\textwidth]{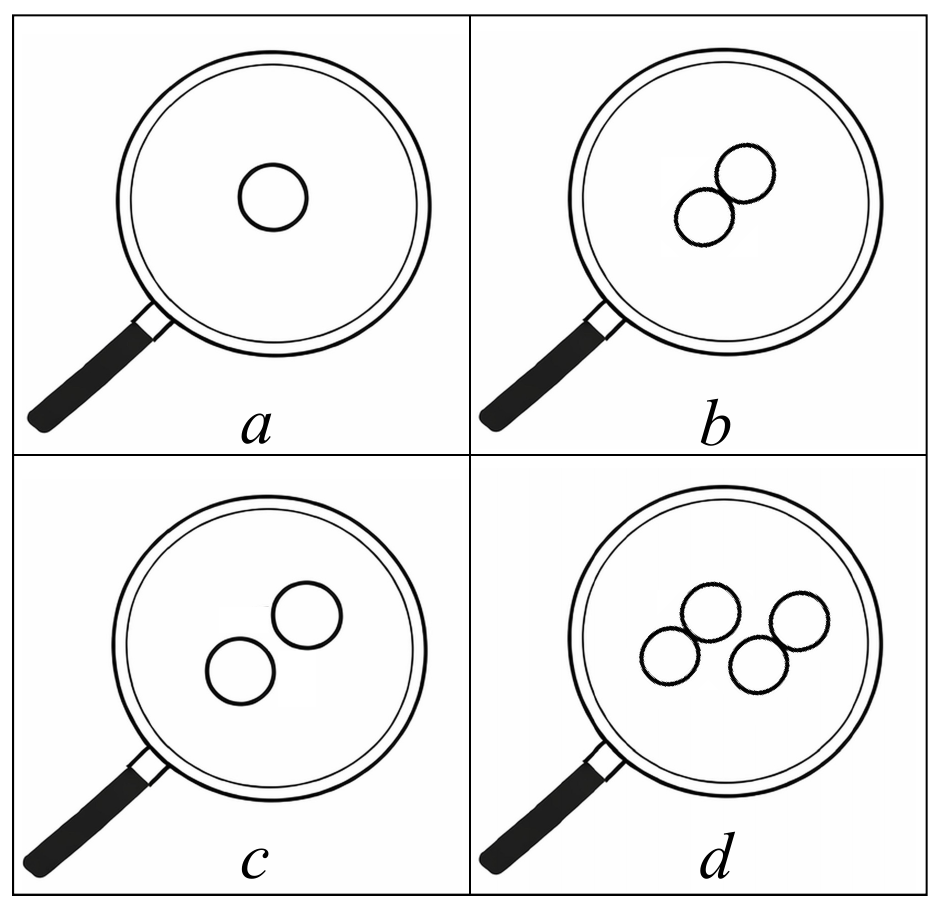} %
    \caption{Schematic diagrams of gluon distributions at different density scales.
}\label{fig:2}

\end{figure}

Let us consider the following states of gluons within a proton, each characterised by a different density, prior to detection: (a) a single gluon, (b) a dipole, (c) a pair of closely spaced gluons and (d) a pair of closely spaced dipoles (Fig.~\ref{fig:2}). In the parton picture, the evolution equations derived from the renormalisation group can be described as the variation in the Feynman amplitude following the addition of an initial parton~\citep{Altarelli:1977zs}.

\begin{figure}[htbp]
  %\centering
  \includegraphics[width=.91\textwidth]{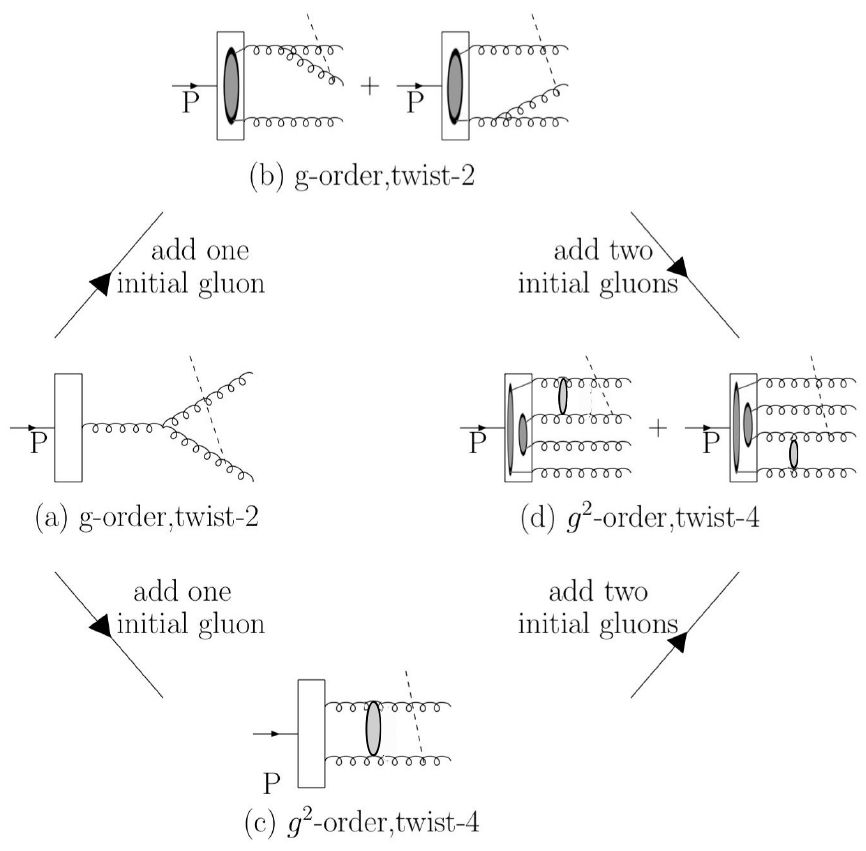}
  \caption{Evolution of the evolution equations beginning with the DGLAP equation (\textbf{a}) and they lead to (\textbf{b}) the BFKL equation, (\textbf{c}) the GLR-MQ-ZRS equation and (\textbf{d}) the ZSR equation. The dashed line denotes a virtual current probing the gluon distribution. Note that the four evolution equations form a closed circuit, which implies consistency among the four evolution equations.
}\label{fig:3}
\end{figure}

\begin{figure}[htbp]
  %\centering
  \includegraphics[width=1\textwidth]{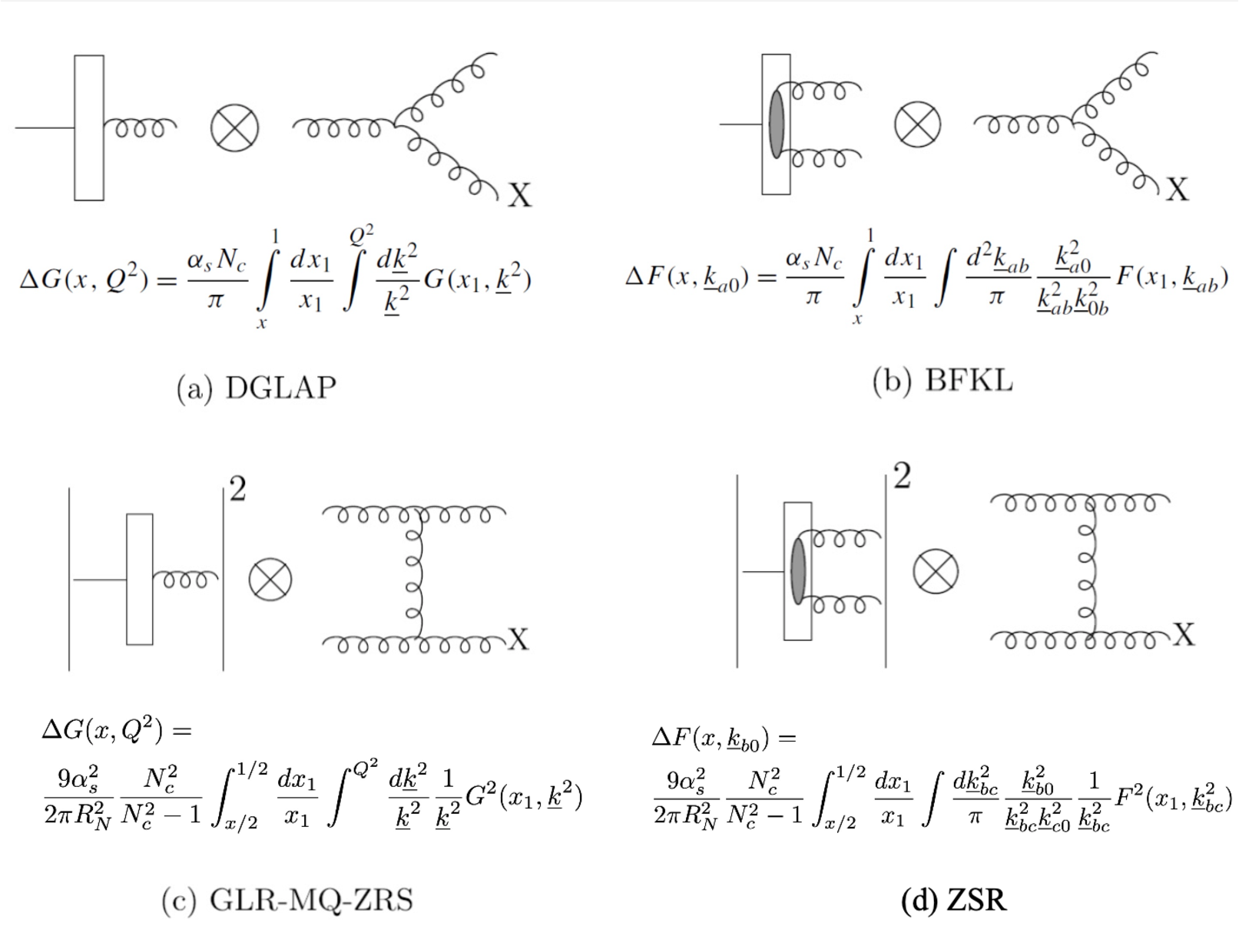}
  \caption{The elemental amplitudes for the four evolution equations based on Fig.~\ref{fig:3}.
}\label{fig:4}
\end{figure}

The simplest DGLAP equation (Fig.~\ref{fig:3}a) neglects the correlations between particles in the initial state; it exhibits a linear dependence on the initial gluon distribution. However, in regions of high particle density, this assumption no longer holds, as the wave functions of the gluon begin to overlap spatially. We sequentially add correlated initial gluons to the lowest-energy amplitude in Fig.~\ref{fig:3}a. The gluon distribution observed by the gluon probe (the dashed line in Fig.~\ref{fig:3}) via fundamental QCD interactions (gluon splitting and recombination) is described by the fundamental amplitudes in Fig.~\ref{fig:4}, which constitute the four QCD evolution equations for different approximations.

It is worth noting that these amplitudes together form a closed loop. These fundamental amplitudes constitute the basic building blocks of the four evolutionary equations mentioned above. This symmetric structure indicates that the corresponding system of equations is self-consistent. This reciprocity in the structure is not a mere mathematical coincidence.  Fig.~\ref{fig:4} reveals the correlations among these equations: the DGLAP equation and the BFKL equation (or the GLR-MQ-ZRS equation and the ZSR equation) share the same evolutionary dynamics, i.e., gluon splitting (or gluon recombination). Using this reciprocity, we have the following analogy in Fig.~\ref{fig:4}.

\vspace{-6pt}

\begin{equation}\label{eq:1}
\frac{\mathrm{d}\underline{k}^2}{\underline{k}^2}\longleftrightarrow
\frac{\mathrm{d}^2\underline{k}_{ab}}{\pi}\frac{\underline{k}^2_{a0}}{\underline{k}^2_{ab}\underline{k}^2_{0b}}
\end{equation}

\begin{equation}\label{eq:2}
G(x,\underline{k}^2)\longleftrightarrow
F(x,\underline{k}_{ab}).
\end{equation}

{{From a broader perspective, the closed circuit displayed in Figs.~\ref{fig:2} and \ref{fig:3} may be interpreted as a generalized structural symmetry in the space of QCD evolution equations. This symmetry is not a conventional Noether symmetry acting on space-time variables or internal quantum numbers; rather, it acts on the effective descriptions themselves. Each vertex of the circuit represents an evolution datum, consisting of a distribution function, an evolution kernel and the corresponding phase-space measure. The evolution equation then plays the role of a morphism that maps a given distribution into its scale or rapidity variation. The relations among different evolution equations, such as the correspondence between the integrated gluon distribution $G(x,\underline{k}^2)$ and the unintegrated distribution $F(x,\underline{k}_{bc})$, may be viewed as higher-level morphisms connecting different evolution morphisms. In this sense, the closed circuit expresses a categorical reciprocity: lifting a linear collinear evolution to a small-x transverse-momentum evolution and then introducing nonlinear recombination should lead to the same structural sector as first introducing nonlinear recombination and then performing the corresponding small-x lift. Schematically, this can be written as
\[
\begin{tikzpicture}
    \node at (0,0) (A) {${\text{DGLAP}}$};
    \node at (3,0) (B) {${\text{BFKL}}$};
    \node at (0,-2) (C) {${\text{GLR/MQ/ZRS}}$};
    \node at (3,-2) (D) {${\text{ZSR}}$};
    
    \draw[->] (A) -- node[above] {$R$} (B);    
    \draw[->] (A) -- node[left] {$N$} (C);     
    \draw[->] (B) -- node[right] {$N$} (D);    
    \draw[->] (C) -- node[below] {$R$} (D);    
\end{tikzpicture}
\]
or equivalently, $N \bigotimes  R\simeq R\bigotimes N$, where $R$ denotes the change of representation from the collinear gluon distribution to the unintegrated or dipole distribution, and $N$ denotes the nonlinear extension associated with gluon recombination. The ZSR equation is therefore not an isolated construction, but the missing corner required by the closure of this generalized symmetry structure. We use the term generalized structural symmetry in an analogical sense~\citep{Gaiotto:2014kfa, Sharpe:2015mja}. It should not be confused with an exact Noether symmetry or with a rigorously established higher-form symmetry of QCD. Rather, it denotes a reciprocity structure among different effective evolution descriptions, in which distribution functions and evolution kernels play the role of objects, evolution equations play the role of morphisms, and the correspondences among different evolution equations play the role of higher-level morphisms.}}

Thus, provided that three of the evolution kernels are known, the real part of the fourth equation can be derived directly~\citep{Zhu:2016qif}. Once the most basic amplitudes are known, the standard procedure in quantum field theory is to sum over the contributions from all relevant processes~\citep{Sterman1993}. This is particularly crucial for evolution equations with infrared singularities. However, it is precisely this step that has posed obstacles in the study of nonlinear evolution equations; many authors have adopted the approach of removing the singular terms, thereby losing the opportunity to discover GC. Work~\citep{Zhu:1998hg} overcame these challenges using TOPT and constructed new quantum chromodynamics evolution equations. Based on a complete derivation from quantum field theory, the results are in full agreement with the aforementioned conjecture. See~\citep{Zhu:2016qif} for details. The resulting evolution equation, denoted as the ZSR equation at small x in the leading approximation, reads

\begin{equation}\label{eq:3}
\begin{split}
-x\frac{\partial F(x, \underline{k}^2)}{\partial x}\simeq
&\frac{3\alpha_s \underline{k}^2}{\pi}\int^\infty_{\underline{k}^2_0} \frac{\mathrm{d}\underline{k}'^2}{\underline{k}'^2}
\left\lbrace
\frac{F(x,\underline{k}'^2) - F(x, \underline{k}^2)}{\vert \underline{k}'^2-\underline{k}^2\vert} + \frac{F(x, \underline{k}^2)}{\sqrt{\underline{k}^4 + 4\underline{k}'^4}}
\right\rbrace\\
&-\frac{81}{16}\frac{\alpha^2_s}{\pi R^2_N}\int^\infty_{\underline{k}^2_0}
\frac{\mathrm{d}\underline{k}'^2}{\underline{k}'^2}
\left\lbrace
\frac{\underline{k}^2F^2(x, \underline{k}'^2) - \underline{k}'^2 F^2(x, \underline{k}^2)}{\underline{k}'^2\vert \underline{k}'^2 - \underline{k}^2\vert}
+\frac{F^2(x, \underline{k}^2)}{\sqrt{\underline{k}^4 + 4\underline{k}'^4}}
\right\rbrace,
\end{split}
\end{equation}

where the linear part is given by the BFKL contribution, and we take the cylindrically symmetric approximation. When the interference terms are neglected, the ZSR equation naturally reduces to the GLR equation, and even to the BK equation~\citep{Ruan:2006ud}; and reduces to the BFKL equation when the non-linear terms are neglected. Since the first three equations (DGLAP, BFKL and GLR-MQ-ZRS)
have been widely used in experimental research and are regarded as crucial support for perturbative QCD theory within the Standard Model, this has aroused our interest in the fourth equation mentioned above, particularly as it predicts a previously unobserved GC.

\section{The Origin of GC\label{sec3}}
In quantum mechanics, multiple bosons can occupy the same quantum state and cannot be distinguished from one another.
This phenomenon is collectively referred to as quantum condensation, or simply condensation~\citep{1995}. How can boson condensation be achieved? What properties does the condensed matter possess? These have long been subjects of interest across multiple disciplines.
A novel form of boson condensation, GC, appears in the solutions of Equation~(\ref{eq:3}); it can be directly tested in next-generation hadron colliders, and at present can also be observed indirectly in the energy spectra of cosmic-rays.

This GC solution of Equation~(\ref{eq:3}) represents a butterfly effect triggered by chaos in protons. Chaos is a non-linear
phenomenon that is ubiquitous in nature, and it frequently arises in non-linear iterative equations~\citep{Feldman2019}. The QCD evolution equation is a differential-integral equation solved iteratively. Consequently, the search for chaos in non-linear QCD evolution equations has long been a subject of considerable interest. Interestingly, Equation~(\ref{eq:3}) exhibits chaotic behaviour with a positive Lyapunov exponent~\citep{Steven2024}. Numerical calculations indicate that this is due to the presence of so-called Lipatov singularities in the non-linear terms of the equation~\citep{Lipatov:1976zz}.

In high-energy hadronic processes, the transverse momentum of gluons varies arbitrarily with gluon splitting and fusion, giving rise to the so-called Lipatov singularity. According to standard quantum field theory, this infrared singularity should be regularised by summing over all relevant Feynman diagrams, ultimately yielding Equation~(\ref{eq:3}), in which both the linear and non-linear terms contain the regularised Lipatov singularity. The non-linear term in Equation~(\ref{eq:3}) transforms the weak jumps in the gluon distribution into intense chaotic oscillations. The standard criterion for chaos is that the system possesses a positive Lyapunov exponent, indicating that the system is highly sensitive to minute variations in initial conditions. Calculations confirm that chaos can emerge in QCD evolution equations. However, chaotic effects alone cannot generate the GC solution in Equation~(\ref{eq:3}). The regularised non-linear kernel plays a second crucial role, here transforming chaotic oscillations into strong shadowing and antishadowing effects, the latter of which ultimately gives rise to the GC phenomenon. To elaborate, in Equation~(\ref{eq:3}) we have~\citep{Wang:2023zop, Zhu:2017ydo, Zhu:2022bxi}

\begin{equation}\label{eq:4}
\left[\frac{\underline{k}^2F^2(x,\underline{k}'^2)}{\underline{k}'^2\vert
\underline{k}'^2-\underline{k}^2\vert}-\frac{\underline{k}'^2F^2(x,\underline{k}^2)}{\underline{k}'^2\vert
\underline{k}'^2-\underline{k}^2\vert}\right]_{\underline{k'}^2\sim\underline{k}^2}\subseteq
\frac{\mathrm{d}}{\mathrm{d}\underline{k}'^2}\left[\frac{\underline{k}^2}
{\underline{k}'^2} F^2(x,\underline{k}'^2)\right]
\end{equation}

Once chaos arises, rapid oscillations in gluon density will induce both negative and positive non-linear corrections to
$F(x, \underline{k}^2)$ via the derivative operation. The former gradually dampens the growth of gluon density as $F$ is simultaneously suppressed, whilst the latter, acting as a positive feedback process, accelerates and intensifies in proportion to the increasing $F$.
These alternating shadowing and antishadowing effects produce a synergistic effect near the critical momentum, driving the aggregation of gluons and thereby forming GC.

Equation~(\ref{eq:3}) neglects higher-order corrections. A key question is: once higher-order corrections are taken into account, will the chaotic effects disappear from the evolution equation? This issue has been discussed in~\citep{Zhu:2017ydo, Zhu:2022bxi}, and we summarise their conclusions as follows: the GC effect arises from the singular non-linear evolution kernel and local momentum conservation. The multiple singular structures generated by higher-order corrections should be able to be cancelled out by contributions from Feynman diagrams of the same order. The resulting non-linear evolution equation still satisfies the conditions required for GC to occur. Although higher-order QCD corrections may alter the value of $k_c$, the simplified form of the GC solution contains only a few parameters $(x_c,k_c)$, and these parameters can be determined from experimental data.

Another point to be discussed is the substitution of the square of the single-particle (or single-dipole) distribution for the two-particle (or two-dipoles) correlation function  in the derivation of the equation. We have found that introducing a correlation function in place of the above simplified model delays the GC threshold but does not eliminate the GC phenomenon. These uncertainties, along with other uncertainties in the equations (such as higher-order corrections) are effectively absorbed into the small number of free parameters of the GC model, to be determined by observation. It must be noted that the standard BK equation has no chaotic solutions. However, certain discretised or modified BK-type mappings can exhibit chaos. Even so, such BK equations cannot give rise to GC, as they lack the mechanisms necessary to form strong antishadowing.

\section{Looking for GC\label{sec4}}
GC causes significant changes in the gluon distribution function, which should be observable in proton colliders. The fact that this remains undetected can be attributed to the insufficient energy of the LHC, meaning that the peak at which GC occurs has not yet entered the interaction region. We therefore turn our attention to cosmic-rays, where high-energy gamma rays have two possible production mechanisms: the leptonic and the hadronic scenarios~\citep{Aharonian_2004}. The latter refers to $pp(A)\rightarrow \pi^0$ and $\pi^0\rightarrow 2\gamma$. In $pp(A)$ collisions, approximately half of the energy from the parent proton is absorbed by valence quarks, which subsequently form leading particles. The remaining half is converted into secondary hadrons, primarily pions, in the central region. The energy spectrum formula for these photons in the laboratory system is

\begin{equation}\label{eq:5}
\Phi_{\gamma}(E_{\gamma})=C_{\gamma}\underbrace{\left(\frac{E_{\gamma}}{{\rm GeV}}\right)^{-\beta_{\gamma}}}_{(1)}
\int_{E_{\pi}^{min}}^{E_{\pi}^{max}}\mathrm{d}E_{\pi}\underbrace{
\left(\frac{E_p}{{\rm GeV}}\right)^{-\beta_p}}_{(2)}\underbrace{N_{\pi}(E_p,E_{\pi})}_{(3)}\underbrace{
\frac{\mathrm{d}\omega_{\pi-\gamma}(E_{\pi},E_{\gamma})}{\mathrm{d}E_{\gamma}}}_{(4)},
\end{equation}

where $\Phi_{\gamma}$ is the spectral energy distribution of gamma rays. $C_{\gamma}$ is a normalisation constant that includes the motion factor, (4) represents a typical normalized Quantum Electrodynamics (QED) process.
The physical significance of the other terms on the right-hand side is as follows:

(1) It describes the absorption of gamma rays in the vicinity of the source. For gamma rays originating from extragalactic sources, the absorption of photons by the extragalactic background light (EBL) during long-distance propagation must be taken into account; this will be dealt with separately.

(2) This is the energy spectrum of accelerated protons, which is related to the proton acceleration mechanism. It is generally assumed to follow a power-law (PL) distribution, although a correction factor may apply at the high-energy end of the spectrum.

(3) $N_{\pi}$ is related to the $pp$ collision cross-section. This provides a window for observing the GC effect. Typically, because $N_{\pi}$ involves intractable non-perturbative difficulties, it is replaced by an empirical parameterised expression, which is a curve that rises gradually with collision energy. It does not produce a distinct peak in the gamma spectrum, but shifts the
$\pi^0\rightarrow 2\gamma$ peak at $m_{\pi}/2$ to around 1 GeV, a phenomenon known as the $\pi$-bump. This theoretical prediction has been confirmed by observations~\citep{Fermi-LAT:2013iui}. When the proton collision energy exceeds the range of applicability of the empirical cross-section formula, we must estimate $N_{\pi}$ using the parton model of the proton, thereby providing an opportunity to test GC~\citep{Zhu:2017bvp}. This point is discussed in more detail below.

In high-energy proton collisions, the collision of gluons originating from the two protons gives rise to gluon mini-jets in the central region; these mini-jets rapidly hadronise into secondary particles, which are primarily
pions. The multiplicity $N_g$ of the gluon mini-jets is related to the gluon distribution $F(x,k)$. Converting $N_g$ to the pion multiplicity $N_{\pi}$ in the formula presents the problem of gluon mini-jet hadronisation, for which no solution currently exists.
Collision heating is a relatively slow process; thermal equilibrium is only reached after numerous random collisions have dissipated the relative momentum. The massive number of condensing gluons rapidly injected into the central region of the collision can absorb sufficient collision energy to greatly increase the pion production rate, to the point of saturation that is, almost all available collision energy is used to produce new particles via gluons.  Although this is a reasonable conjecture, it lacks rigorous theoretical proof. Can this important assumption be confirmed by observation? The answer is yes. We shall discuss this later. For now, it suffices to note that, utilising the pion saturation assumption and relying solely on the two universal principles of energy conservation and relativistic covariance, one can derive a spectral distribution exhibiting GC characteristics, namely the typical broken power-law (BPL)~\citep{Zhu:2017bvp}.

\begin{equation}\label{eq:6}
E_{\gamma}^2\Phi^{GC}_{\gamma}(E_{\gamma})\simeq\left\{
\begin{array}{ll}
\frac{2e^bC_{\gamma}}{2\beta_p-1}(E_{\pi}^{GC})^3\left(\frac{E_{\gamma}}{E_{\pi}^{GC}}\right)^{-\beta_{\gamma}+2} \\ {\rm ~~~~~~~~~~~~~~~~~~~~~~~~if~}E_{\gamma}\leq E_{\pi}^{GC},\\\\
\frac{2e^bC_{\gamma}}{2\beta_p-1}(E_{\pi}^{GC})^3\left(\frac{E_{\gamma}}{E_{\pi}^{GC}}\right)^{-\beta_{\gamma}-2\beta_p+3}
\\ {\rm~~~~~~~~~~~~~~~~~~~~~~~~ if~} E_{\pi}^{GC}<E_{\gamma}<E_{\pi}^{cut},\\\\
\frac{2e^bC_{\gamma}}{2\beta_p-1}(E_{\pi}^{GC})^3\left(\frac{E_{\gamma}}{E_{\pi}^{GC}}\right)^{-\beta_{\gamma}-2\beta_p+3}
\exp\left(-\frac{E_{\gamma}}{E_{\pi}^{cut}}+1\right).
\\ {\rm~~~~~~~~~~~~~~~~~~~~~~~~ if~} E_{\gamma}\geq E_{\pi}^{cut},
\end{array} \right.
\end{equation}

Besides, two values of $E_{\pi}^{GC}$ and $E_{\pi}^{cut}$ obey

\begin{equation}\label{eq:7}
E_{\pi}^{cut}=e^{b-a}\sqrt{\frac{2m_p/{\rm GeV}}{k_c^2}}(E_{\pi}^{GC})^2
\end{equation}

where $a\equiv 0.5\ln (2m_p/{\rm GeV})-\ln (m_{\pi}/{\rm GeV})+\ln k$
and $b\equiv \ln (2m_p/{\rm GeV})-2\ln (m_{\pi}/{\rm GeV})$ $+\ln k$.

Another useful formula in the GC model is

\begin{equation}\label{eq:8}
E_p=\frac{2m_p}{m^2_{\pi}}E_{\pi}^2,
\end{equation}

which gives the energy of the incident proton
corresponding to $E_{\pi}$ in the laboratory frame. Equation~(\ref{eq:6}) is known as the GC spectrum. In the literature on cosmic gamma rays, similar BPL distributions often appear as empirical formulas used to fit observational data. The GC spectrum provides a simple and reasonable explanation for this. We will now discuss the reliability of several assumptions in the GC spectra (\ref{eq:6}). To determine the validity of this equation, one must examine the assumptions it incorporates and its dependence on those assumptions. The former include photon absorption, the proton energy spectrum, and the fact that $N_{\pi}$ must follow the PL; all are indispensable. This is because any deviation from the PL in any of these factors would distort Equation~(\ref{eq:6}). We therefore focus on the PL solution for $N_{\pi}$.

We use the sensitivity of parameter $e^b$ to $m_{\pi}$ in the GC spectrum, Equation~(\ref{eq:6}), to test the validity of the pion saturation assumption by examining the data.  Assuming pions in the centre of the collision have transverse momentum with an average value of $\langle p_T\rangle$. We regard these pions as virtual pions with an effective mass of
$m^*_{\pi}=\langle p_T\rangle^2/(2m_{\pi})$. By repeating the calculations in Equation~(\ref{eq:6}), we obtain the value of $b^*$ corresponding to $\langle p_T\rangle\neq 0$.
Since the actual transverse momentum of a pion varies over the interval 0$\sim$$\langle p_T\rangle$, the dependence of the number of particles on transverse momentum will deviate significantly from the PL distribution. The width of this deviation is determined by $m_{\pi}^*$.

In $pp$ collisions at the LHC, the freezing temperature $T_f$~$\simeq$ 100$\sim$150 $\mathrm{MeV}$ is the temperature at which the system ceases to interact. If only pure thermal motion (without collective flow) is considered, the average transverse momentum of pions can be calculated from the thermal distribution. Assuming the transverse momentum spectrum follows an exponential form $dN/dp_T\propto p_T \exp(- p_T/T_f)$,
the average transverse momentum in the purely thermal regime corresponding to a freezing temperature of 100$\sim$150\ $\mathrm{MeV}$ is approximately 0.2$\sim$0.35 $\mathrm{GeV}$. Let us now consider a specific example. The Large Hadron Astrophysics Observatory (LHAASO) has detected the energy spectrum of the microquasar SS 433 extending up to ultra-high energies of 100 TeV~\citep{LHAASO:2024psv}.
This cannot be explained by a single lepton component alone. In the ultra-high-energy range, its radiation is spatially coincident with giant atomic clouds (a condition required for a hadronic origin) yet it still cannot be fitted by a conventional hadronic scenario.

\begin{figure}[htbp]
  %\centering
  \includegraphics[width=1\textwidth]{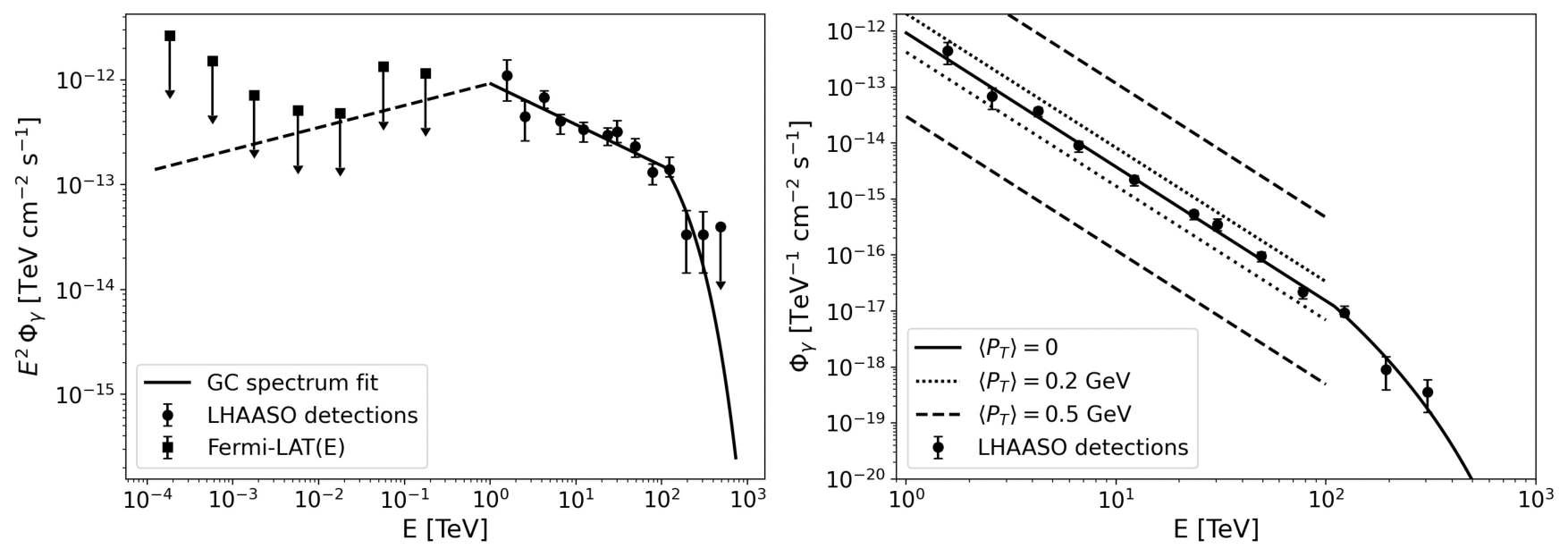}
  \caption{Spectral energy distribution (SED) of SS 433 measured by LHAASO~\citep{LHAASO:2024psv} and comparison with the GC spectra (solid curve).
   $C_{\gamma}=5.33\times 10^{-15}(\mathrm{GeV}^{-2}\mathrm{cm}^{-2}\mathrm{s}^{-1})$, $\beta_{\gamma}=1.79$, $\beta_p=0.80$, $E_{\pi}^{GC}=1.0\ \mathrm{TeV}$ and
   $E_{\pi}^{cut}=110\ \mathrm{TeV}$. Different values of the average transverse momentum of pions allow the regions where the observation points deviate from the PL to be delineated by different lines, where $\langle p_T\rangle=0.2\ \mathrm{GeV}$ corresponds to the frozen region, i.e. the condition required by the GC spectrum: the average transverse momentum of pions can be neglected, and the pion yield shows~saturation.
    }\label{fig:5}
\end{figure}

Using the GC spectra, we have successfully interpreted the energy spectrum of SS 433 with just five free parameters. To test the pion saturation assumption, Fig.~\ref{fig:5} uses solid and dotted lines to delineate the possible regions where the PL deviates from that predicted by the saturation assumption when $\langle P_T\rangle\neq 0$ is present; Fig.~\ref{fig:5} also shows the spectrum observed by Fermi-LAT, with the overall spectrum exhibiting a typical BPL. The above analysis shows that the gamma-ray spectrum is highly sensitive to the saturation of pions in the central region. The transverse momentum of pion causes the gamma-ray spectrum to deviate significantly from the PL. This allows us to infer indirectly the presence of the GC effect.

Interestingly, the mean transverse momentum of 200 MeV also corresponds to the pure thermal mean transverse momentum at the freeze-out temperature of $AA$ collisions.
The hot, dense medium produced by $AA$ collisions (such as quark--gluon plasma or hadronic gas) undergoes expansion and cooling until it reaches a stage where elastic and inelastic collisions between particles have essentially ceased. At this point, the momentum distribution of the particles is \textit{frozen.} After freezing, the mean free path of the particles exceeds the size of the system; the particles flow freely, and their momentum distribution no longer changes, retaining the thermodynamic and kinetic characteristics present at the time of freezing until they are recorded by the detector. The pure thermal mean transverse momentum corresponding to a freeze-out temperature of 100$\sim$150 MeV is approximately 0.2$\sim$0.35~GeV, which corresponds precisely to the freezing temperature. This scale is a concept in heavy-ion collisions. Currently, in ultra-high-energy collisions in small systems, the $pp(A)$ initial state may also exhibit freezing. However, this is attributed to the saturation of $\pi$-meson production, rather than to the expansion of the final-state system.

High-energy $p-p$ collisions are an indispensable physical process in the universe, occurring extensively in extreme astronomical environments such as supernova explosions
and their remnants, stellar collapse, neutron star mergers, and black hole formation. Since gluon condensation can be triggered whenever the energy of proton-proton collisions is sufficiently high, resulting in gamma-ray spectra with GC features. Consequently, we can also detect traces of GC in a wide range of other ultra-high-energy gamma-ray sources; six observational examples of the GC spectra
are listed below.

{\bf{J1534-571}~\citep{HESS:2016azh, Araya:2017cxq}}

\begin{figure}[htbp]
  %\centering
  \hspace{-21pt}\includegraphics[width=1\textwidth]{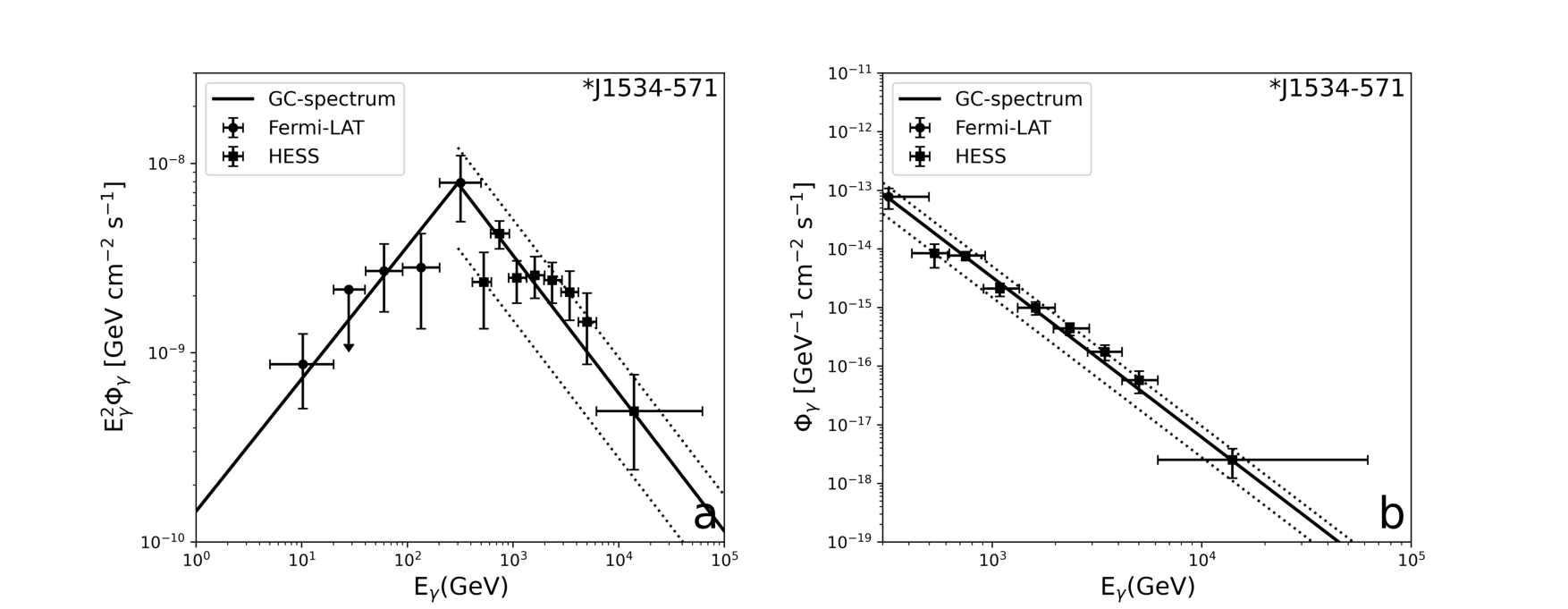}
  \caption{SED of J1534-571~\citep{HESS:2016azh, Araya:2017cxq} and comparison with the GC spectra (solid curve). Parameters are
$C_{\gamma}=3.98\times 10^{-18}(\mathrm{GeV^{-2}cm^{-2}s^{-1})}$, $\beta_{\gamma}=1.3, \beta_p=1.2, E_{\pi}^{GC}=300\ \mathrm{GeV}$ and
$\chi^2/\mathrm{d.o.f.}=2.0$. The area between two point lines is the frozen range.
    }\label{fig:6}
\end{figure}

HESS J1534-571 is a very-high-energy (VHE) gamma-ray source discovered by the HESS observatory that displays a very hard power-law
spectrum. Because of the absence of a ``$\pi$-bump'', although no convincing X-ray counterpart has been found, it remains classified as a leptonic scenario. In the existing literature, the leptonic scenario was introduced primarily to accommodate the overall broadband SED, but not to examine, on the same scale and with the same quantitative precision, the local GeV–TeV power-law segment displayed in Fig.~\ref{fig:6}b. This distinction is important, because the physical diagnostic value of the spectrum lies precisely in that clean power-law behavior. At present, no secure nonthermal X-ray synchrotron emission has been detected from this source, so the synchrotron branch required by the leptonic interpretation is not independently measured, but effectively inferred through parameter choices that render the predicted X-ray flux consistent with the available upper limits. In this respect, the leptonic solution is better understood as an underconstrained consistency fit than as a broadband explanation directly supported by multi-wavelength evidence. By contrast, the GC interpretation is constructed to account specifically for the observed GeV–TeV spectral morphology itself.
We use Equation~(\ref{eq:6}) to fit J1534-571 in Fig.~\ref{fig:6}. In terms of goodness of fit (reduced chi-square), the GC model clearly has an advantage over the lepton model, as the latter contains a large number of uncertain free parameters in the low-energy spectrum, which significantly reduces the goodness of fit. The BPL formed by the HESS and Fermi-Lat data provides a characteristic signature of the GC spectrum, so we judge that the source may belong to the hadronic scenario. In particular, it naturally targets the nearly ideal power-law structure shown in Fig. 6b, which is the very feature that is not explicitly tested in previous leptonic modeling. We therefore suggest that the present data do not favor a leptonic origin in any decisive sense, but instead leave open the possibility that HESS J1534-571 contains a hadronic component whose spectral signature is more naturally captured in the GC framework.

{\bf{J1825-137}~\citep{Principe:2020zqe}}

\begin{figure}[htbp]
  %\centering
 \includegraphics[width=.931\textwidth]{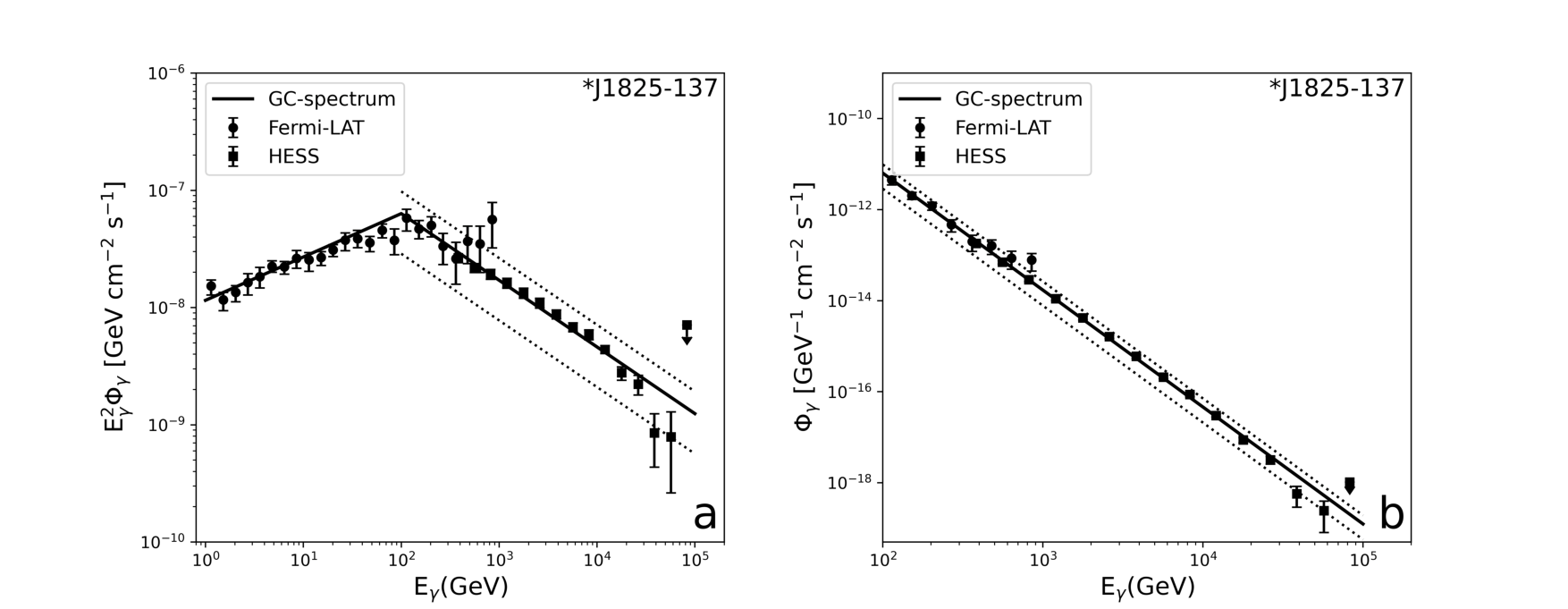}
  \caption{SED of J1825-137~\citep{Principe:2020zqe} and comparison with the GC spectra (solid curve). Parameters are
$C_{\gamma}=5.69\times 10^{-16}(\mathrm{GeV^{-2}cm^{-2}s^{-1}})$, $\beta_{\gamma}=1.63, \beta_p=0.97, E_{\pi}^{GC}=100\ \mathrm{GeV}$ and
$\chi^2/\mathrm{d.o.f.}=3.77$. The area between two point lines is the frozen range.
    }\label{fig:7}
\end{figure}

HESS J1825-137 is the largest known gamma-ray pulsar wind nebula (PWN). J1825-137
has been analysed using more than 11 years of Fermi-LAT data combined with the results of the HESS. It is classified as a leptonic scenario, but lacks a convincing broadband counterpart. Work~\citep{Principe:2020zqe} points out that the leptonic model is not energetically adequate for the observed X-ray and gamma-ray emission, even when supplemented with conventional hadronic components. The spectral pattern shown
in Fig.~\ref{fig:7} is similar to that of the GC-spectrum, and a possible explanation can be provided by considering the J1825-137 spectrum as a GC-spectrum derived from SNR. Our purpose is not to argue against the established PWN picture, nor to claim that a leptonic origin should be excluded. Rather, we wish to emphasize that the existing studies are primarily focused on the source morphology and transport physics, whereas the GC fit presented in Fig.~\ref{fig:7} addresses a different question, namely whether the observed spectral shape itself can also be described in a compact hadronic framework with a GC-type broken-power-law form. In this sense, the GC interpretation should be regarded as a supplementary, spectrum-oriented description that may help highlight an additional hadronic possibility, without replacing the source’s well-developed PWN identification. 

{\bf{J1834-087}~\citep{HESS:2014bhh}}

\begin{figure}[htbp]
  %\centering
  \includegraphics[width=.931\textwidth]{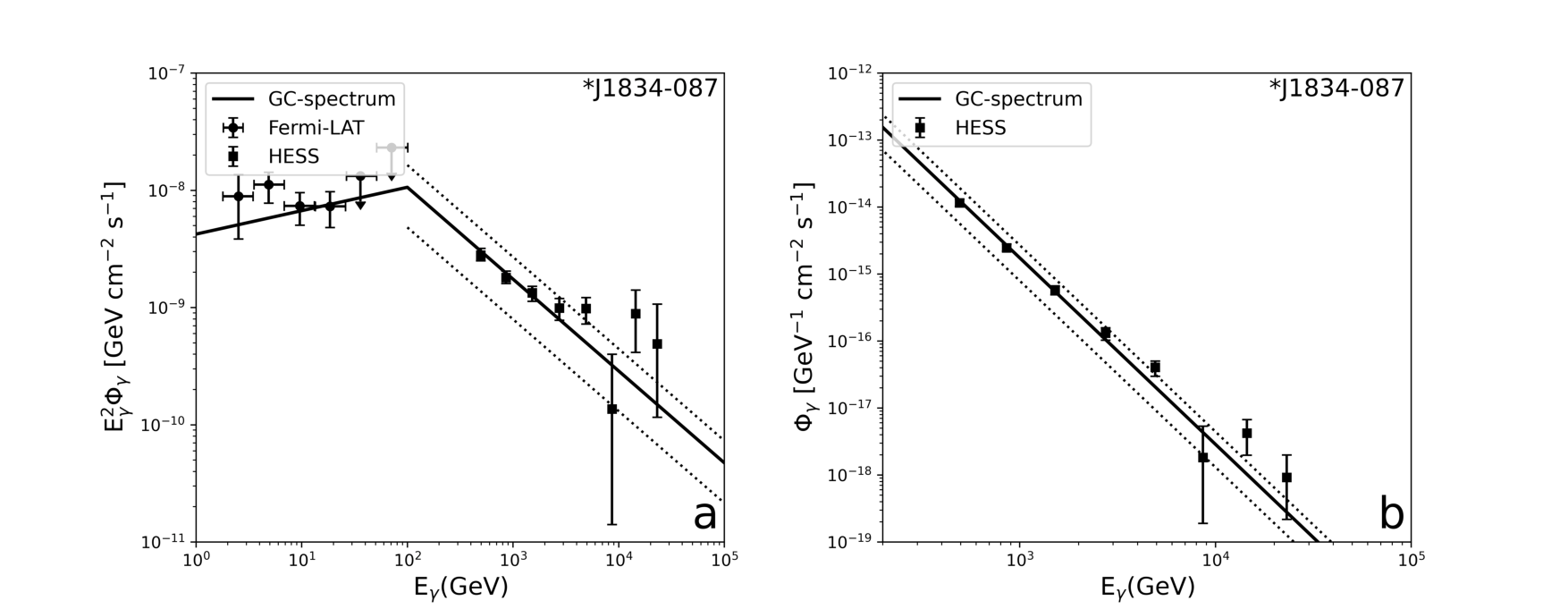}
  \caption{SED of J1834-087~\citep{HESS:2014bhh} and comparison with the GC spectra (solid curve). Parameters are
$C_{\gamma}= 10^{-16}(\mathrm{GeV^{-2}cm^{-2}s^{-1}}), \beta_{\gamma}=1.8, \beta_p=0.99, E_{\pi}^{GC}=100\ \mathrm{GeV}$ and $\chi^2/\mathrm{d.o.f.}=1.54$.
The area between two point lines is the frozen range.
    }\label{fig:8}
\end{figure}

HESS J1834-087 is an unidentified source. Both GeV and TeV source spectra can be described by a power-law form and they show an inflection point around 100 GeV. However, no pulses have been detected in radio, X-ray or high-energy gamma rays. Different scenarios have been explored to explain this gamma-ray emission. Because previous fits remain incomplete, the interpretation is still inconclusive. We find
that it is characterized by the above depiction of the GC-spectrum. The results of the fit are shown in Fig.~\ref{fig:8}. In this context, our purpose is not to claim that the existing interpretations are invalid, but rather to point out that the GC fit in Fig.~\ref{fig:8} addresses the spectral morphology itself in a more direct way. In particular, it provides a compact hadronic description of the smooth GeV–TeV connection and the broken- or quasi-power-law structure, without introducing an additional low-energy electron component or relying on a more complex source decomposition. The GC interpretation should therefore be understood here as an alternative~spectrum.

{\bf{J1912+101}~\citep{HESS:2016azh, Zhang:2019riu}}

\begin{figure}[htbp]
  %\centering
  \includegraphics[width=1\textwidth]{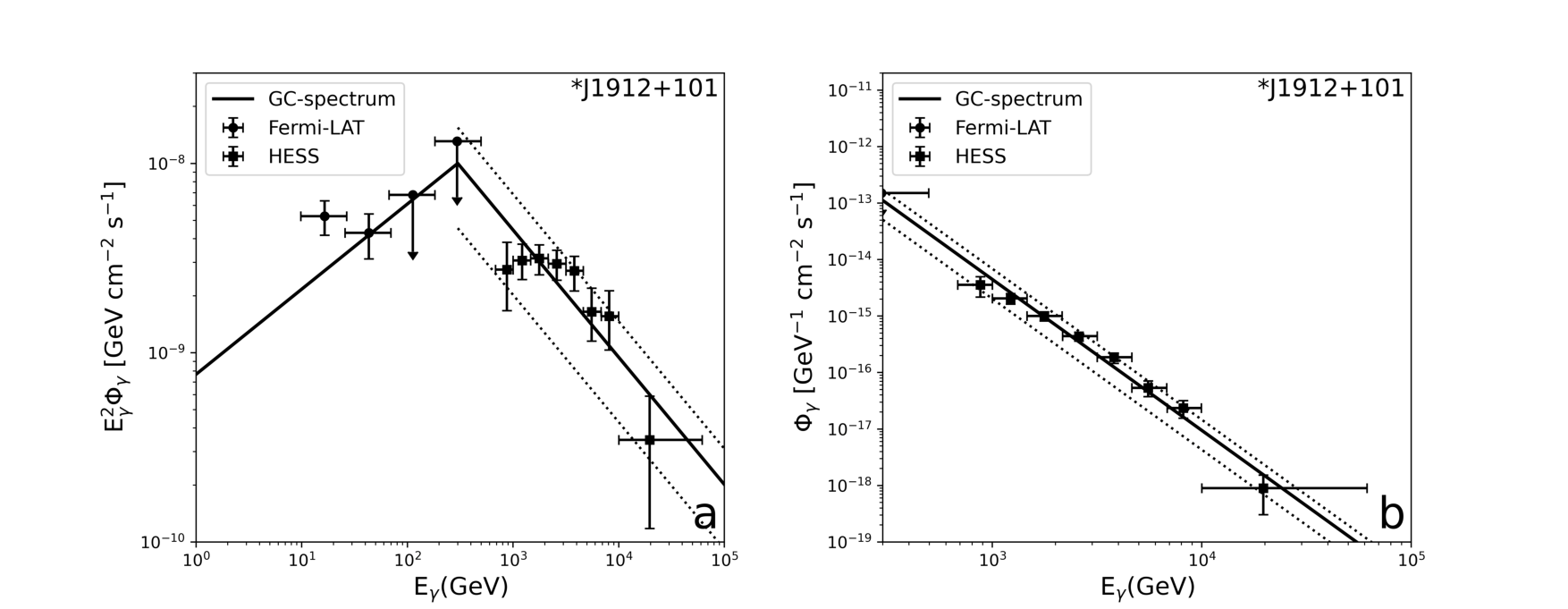}
  \caption{The SED of J1912+101 and comparison with the GC spectra (solid curve). Parameters are
$C_{\gamma}=3.98\times 10^{-18}(\mathrm{GeV^{-2}cm^{-2}s^{-1}}), \beta_{\gamma}=1.55, \beta_p=1.06, E_{\pi}^{GC}=300\ \mathrm{GeV}$ and
$\chi^2/\mathrm{d.o.f.}=1.9$. The area between two point lines is the frozen range.
}\label{fig:9}
\end{figure}

HESS J1912+101 is a GeV-TeV source observed by HESS and Fermi-LAT in a supernova remnant (SNR). HESS J1912+101 appears to be associated with energetic pulsars. VHE observations of these pulsar wind nebulae (PWNe) make it possible to probe the spectrum of accelerated particles and study the associated magnetic field distribution. If true, this source would be the oldest candidate for a VHE-emitting pulsar. High-energy pulsars typically exhibit more compact X-ray nebulae, but no such PWN has been detected in existing X-ray observations. Nevertheless, the emission mechanism of HESS J1912+101 has been hypothesized to belong to the leptonic scenario, and attempts have been made to use it to account for the observed cosmic-ray positron excess. However, since no convincing low-energy (radio and X-ray) counterpart matching the high-energy spectrum has been found, the nature of this source is still uncertain. We do not exclude that the above spectrum is a hadronic spectrum with GC effects. Indeed, the results of the fit to the GC-spectrum are shown in Fig.~\ref{fig:9}. It is clearly a BPL as predicted by the GC-spectrum, both in terms of the fragmentation behavior and the PL shape. Thus, J1912+101 may be dominated by SNRs through hadronic processes, thus limiting their contribution to the positron excess. Under these circumstances, our GC fit should not be read as a definitive rejection of the pulsar-related scenario. Rather, it is intended to show that the observed GeV–TeV spectral shape can also be naturally accommodated within a hadronic framework if the shell/SNR interpretation proves to be correct. In this sense, the GC description offers a spectrum-based hadronic possibility in a source whose broader astrophysical classification remains open.

{\bf{J1844+034}~\citep{HAWC:2023ufk}}

\begin{figure}[htbp]
  %\centering
  \includegraphics[width=1\textwidth]{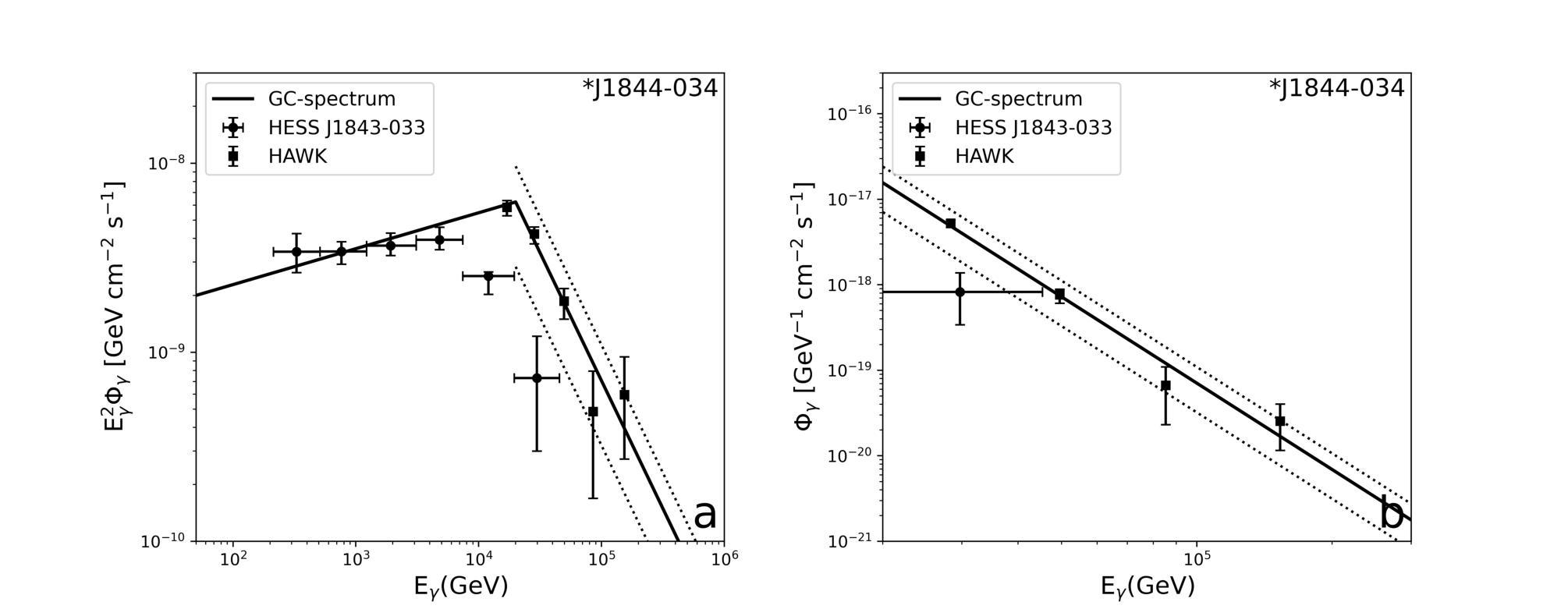}
  \caption{SED of J1844+034 and comparison with the GC-spectra (solid curve). Parameters are
$C_{\gamma}=1.15\times 10^{-23}(\mathrm{GeV^{-2}cm^{-2}s^{-1}}), \beta_{\gamma}=1.81, \beta_p=1.27, E_{\pi}^{GC}=20\ \mathrm{TeV}$ and
$\chi^2/\mathrm{d.o.f.}=0.89$.  The area between two point lines is the frozen range.
}\label{fig:10}
\end{figure}

Although many Galactic TeV gamma-ray sources have counterparts at other wavelengths, there are still several gamma-ray sources that have not been identified so far. J1844-034 is one of such sources of TeV observed by the High Altitude Water Cherenkov Observatory (HAWC)(Fig.~{\ref{fig:10}}). LHAASO has observed a bright extended source, J1843-0338, at a similar location, and, in addition, in 2022, the Tibetan ASGamma experiment released observations of an extended source, TASG J1844-038. It is likely that the observations from these three different multi-TeV gamma-ray observatories all come from the same source. The use of multi-spectral data to fit conventional hadron and lepton particle spectra suggests that both lepton and conventional hadron scenarios are feasible in this source. However, we question the fact that the above fit has neither a connection to the ``$\pi$-bump'' nor a sufficiently large amount of low-energy radiation data matching the TeV energy spectrum. As a new possibility, we use the GC-spectrum to fit the above gamma spectrum. We consider $pp$ collisions, which have a maximum break point $E_{\pi}^{GC}=20\ \mathrm{TeV}$.

The origin of cosmic-rays has long been a central question in astrophysics. Gamma-ray photons from hadronic sources can be used to probe this problem indirectly. However, gamma rays can also be produced by leptonic processes, so careful discrimination is required. We use J1844-034 as an example to discuss this issue. It is worth noting that the hadronic scenarios with and without gluon coalescence effects are judged quite differently on this issue. A remarkable feature of the cosmic-ray spectrum is the ``knee'' phenomenon of about PeV energy scale. One view is that cosmic-rays with energies below the knee are accelerated by objects in the galaxy. There is no accelerator in the Milky Way capable of reaching energies above the PeV scale. Instead, a more general explanation is that there is no limit to the maximum energy that can be reached by galactic accelerators. The conventional hadronic scenario assumes that the proton energy is an order of magnitude larger than the observed gamma energy, i.e., 1.4 PeV. Thus, according to the first viewpoint above, a cut should soon be visible at the end of the diagram. But Equations~(\ref{eq:7}) and (\ref{eq:8}) require $E_p^{max}=10^3\ \mathrm{EeV}$. This seems to support the second view of the ``knee'' phenomenon.

Our understanding is as follows. The Milky Way is only one of countless galaxies in the universe. Therefore, there is
no compelling reason to rule out the possibility of EeVatrons in the Milky Way. It is possible that the Milky Way produces only a small number of extremely energetic protons, making them difficult to detect by terrestrial detectors. Nevertheless, due to the GC effect which efficiently converts proton kinetic energy into radiative energy, the broken-power-law gamma-ray signature produced when EeV protons interact with the ambient medium can be recognized and observed by ground-based observatories.

Future large ground-based gamma-ray telescopes will probe very-high-energy gamma rays in the universe. Among them, the Cherenkov Telescope Array (CTA) (designed to detect gamma rays in the range of 20 GeV to 300 TeV) and the Southern Wide-field Gamma-ray Observatory (SWGO) (targeting gamma rays from 100 GeV to several hundred TeV) will cover the energy ranges of GC models and enable the testing of key predictions of these models. Therefore, we predict that with the establishment of larger ground-based gamma detectors, the no-cut extension of the GC-spectrum beyond the PeV can be observed.

{\bf{B1259-63}~\citep{HESS:2024pfw}}

\begin{figure}[htbp]
  %\centering
  \includegraphics[width=1\textwidth]{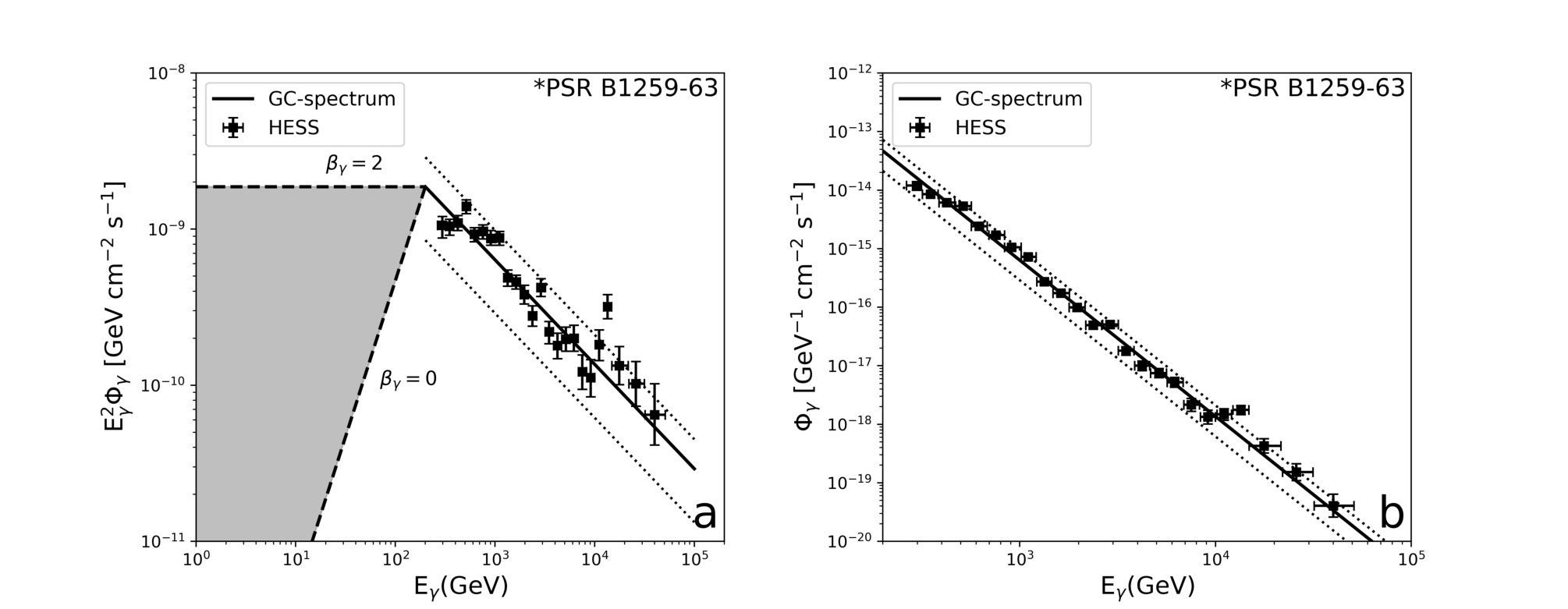}
  \caption{The SED of B1259-63 and comparison with the GC spectra (solid curve). Parameters are
$C_{\gamma}=5.96\times 10^{-18}(\mathrm{GeV^{-2}cm^{-2}s^{-1}}), \beta_p=1.83, E_{\pi}^{GC}=150\ \mathrm{GeV}$ and $\chi^2/\mathrm{d.o.f.}=3.18.$
The area between two point lines is the frozen range.
}\label{fig:11}
\end{figure}

PSR B1259-63 is a binary star system that emits broadband radiation ranging from radio to TeV gamma-rays. The temporal
and spectral properties of this radiation are still poorly understood. In this context, VHE emission is particularly useful when studying radiation processes and particle acceleration in the system. The VHE spectrum obtained from the analysis of the 2021 data set is best described as a power law with a spectral index of $\Gamma=2.65$. However, neither the GeV energy spectrum nor the X-ray signal associated with the TeV energy was found. Even so, this example is categorized as a leptonic scenario, since it is more clearly incompatible with the traditional hadronic scenario.
Surprisingly the PL extends nearly three orders of magnitude and still no cut is visible.
This behavior can be naturally understood within the GC framework. Fig.~\ref{fig:11} is a prediction of Equation~(\ref{eq:6}),
where the shaded area is the possible range of the GC-spectrum before the breakpoint, which corresponds to the $\beta_{\gamma}$ = 0$\sim$2. According to Equation~(\ref{eq:7}), the PL can be extended up to $E_{\pi}^{GC}=315\ \mathrm{TeV}$ if the proton can be accelerated to a sufficiently high energy.

Protons undergoing gluon condensation are a fundamental building block of all matter in the universe; it is therefore reasonable to expect that GC spectra are widely present in various gamma-ray sources. In fact, the GC spectrum has been used to interpret nearly a hundred cosmic gamma-ray spectra, including those from supernova remnants (SNRs), pulsars, active galactic nuclei (AGNs), the galactic centre, and gamma-ray bursts (GRBs).

\section{Discussion}
\begin{figure}[htbp]
  %\centering
  \includegraphics[width=1\textwidth]{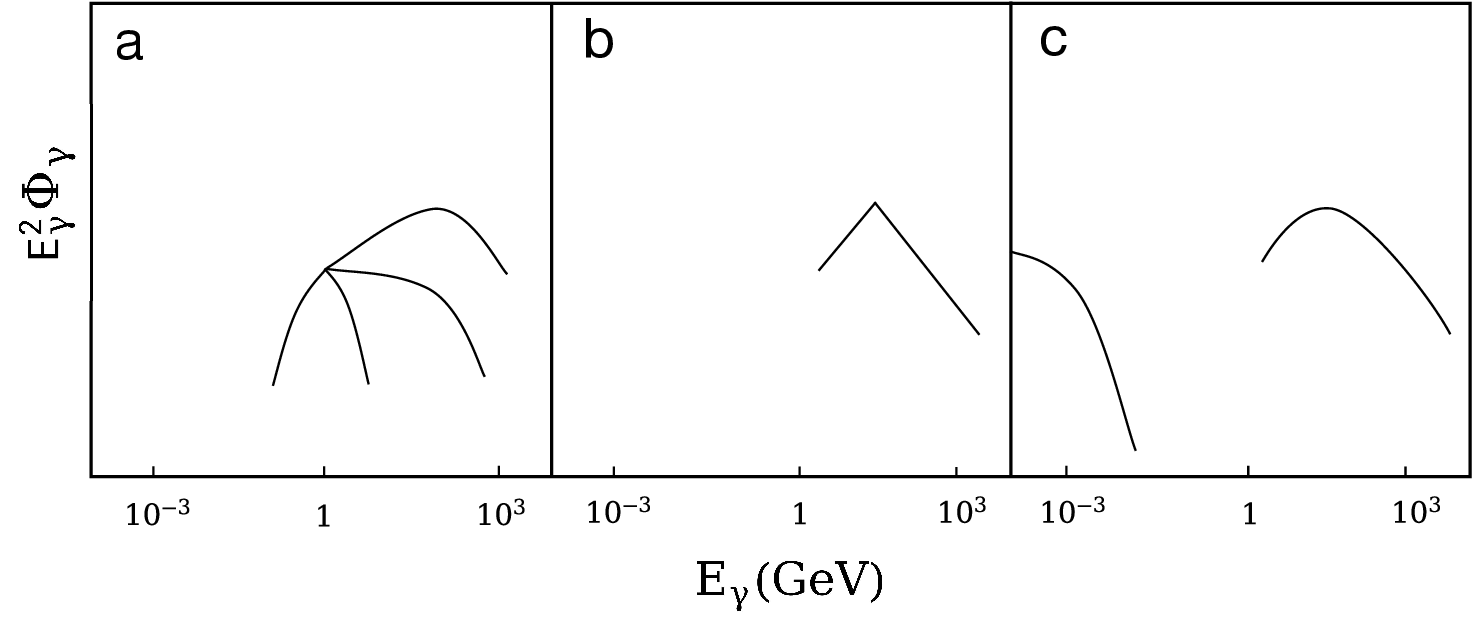}
  \caption{Three types of radiation mechanisms. (\textbf{a}) Hadronic scenario without the GC effect, (\textbf{b})~hadronic scenario with the GC effect and (\textbf{c}) leptonic scenario. (\textbf{a},\textbf{b}) Differ only in the distribution of gluons in the proton, but the shape of the latter's energy spectrum, with its single peak and absence of neutrino radiation, appears more like (\textbf{c}). Moreover, many leptonic candidates lack a convincing low-energy counterpart matched to the high-energy spectrum. It is therefore easy to misclassify case~(\textbf{b}) as case (\textbf{c}).
}\label{fig:12}
\end{figure}

(1) We summarize our thinking in Fig.~\ref{fig:12}, which shows the three VHE gamma-ray spectra discussed in this paper. The hadronic scenario with GC was easily misclassified as a leptonic scenario because it does not have a wide plateau connected to the ``$\pi$-bump''.

The VHE cosmic-gamma radiation is an effective way to observe the structure and evolution of cosmic objects and an important window to study the extreme universe. These emissions are categorized into two types: hadronic scenarios, where protons or nuclei are studied, and leptonic scenarios, where only electronic processes are considered. It is clear that different scenarios depicting the same gamma-ray spectrum will lead to completely different conclusions. A significant fraction of the known VHE gamma-ray sources are still unidentified objects. Also, some of the spectra that have been categorized are considered as leptonic scenarios because they are not related to the ``$\pi$-bump'', but they do not have matching low-energy spectra, casting doubt on their belonging to the leptonic process. GC provides new clues to the above doubts. We find that some VHE gamma-ray events are initially categorized as leptonic events, but are actually more consistent with hadronic events. Thus, when we bring back the GC-spectra, which have been misclassified as leptonic processes, to hadronic processes, we have to revisit our understanding of the structure and evolution of the Universe, which we have gained through VHE gamma-ray observations over the past decades.

(2) GC is a wholly new phenomenon at the deepest level of matter's structure, and its significance extends beyond the realm of particle physics. Firstly, it challenges certain traditional views in astronomy. In the universe, protons are more easily accelerated to high-energy regions than electrons; therefore, the hadronic model should, in theory, be more applicable than the leptonic model in the production of high-energy gamma rays. However, many cases have been classified under the leptonic model because their gamma-ray spectra lack a $\pi$-bump structure. The emergence of GC requires astronomers to re-examine their previous work, particularly research into the search for ultra-high-energy accelerators within the Milky Way.

(3) Pion condensation is a special state of matter in which, under specific extreme conditions, pions condense into a single quantum state. The concept was first proposed by Migdal, who explored the possibility of this phenomenon occurring in nuclear matter in his 1971 paper~\citep{Migdal:1971cu}. Unfortunately, there is currently no conclusive experimental evidence to confirm the existence of meson condensation. Cooling dense and unstable pions is a key challenge in achieving pion condensation. In $p-p$ collisions, a large number of condensed gluons from within the protons flow into the central region, causing a surge in the number of $\pi$ mesons, which are subsequently cooled due to energy conservation constraints. In extreme cases, almost all available collision energy is used to produce pions, briefly forming a dense, low-temperature region in the central zone with the potential to form a $\pi$ meson condensate. This result not only unravels the mystery of $\pi$ meson condensation that has puzzled the physics community for decades, but also reveals new properties of the strong interaction at the microscopic scale.

(4) Comparing GC with Bose--Einstein condensation (BEC), we find that although both involve multiple bosons sharing a single wave function, they represent entirely different physical phenomena. Consequently, GC opens up a new window into understanding nature. For example, the GC model can effectively generate electromagnetic radiation in the universe. Although positron--electron annihilation undoubtedly maximises the efficiency of photon production, this phenomenon is not common in nature. Inverse-Compton scattering is a common mechanism for explaining the VHE gamma-ray spectrum, but it requires an abundance of soft photons to serve as targets for the scattering electrons. GC can be viewed as the manifestation of the butterfly effect in elementary particles. Consequently, there is no need to be concerned about the validity of the perturbative treatment, as once GC exhibits strong chaotic behaviour, it becomes difficult to suppress. Results from fitting the cosmic gamma-ray spectrum to the GC spectrum indicate that %the current LHC p-Pb collision energies are approaching the GC threshold; therefore, it may be possible to observe a weak GC effect at the LHC~\citep{Zhu:2025mcn}, and 
stronger GC signals should be taken into account in the energy range of next-generation hadron colliders. We therefore caution that in the forthcoming Large Hadron Collider programme, further increases in hadron collision energy may lead to the emergence of unexpectedly intense gamma-ray emissions within the accelerator, akin to artificial miniature gamma-ray bursts, which could potentially cause damage to the detectors~\citep{Zhu:2022bxi}.

(5) In nonlinear science, it is not uncommon for nonlinear iterative equations to generate chaos; however, it is a remarkable phenomenon worthy of attention that a single nonlinear evolution equation can simultaneously exhibit robust chaos and masked--unmasked solutions, which further interact to evolve CGC gluons into a condensed state---a phenomenon observable in the high-energy gamma-ray spectrum of the universe.

In summary, this paper reviews the complete logical chain of GC, from its theoretical formulation to potential observational tests. We emphasise that the ZSR equation is not an isolated construct but rather forms, together with the DGLAP, BFKL and GLR-MQ-ZRS equations, an evolutionary system that is interconnected in terms of both dynamics and structure. It is precisely within this framework that a nonlinear evolution kernel preserving the Lyapunov singularity causes the system to exhibit chaotic behaviour with a positive Lyapunov exponent for the first time, driving the gluons to form a condensate near the critical momentum through the synergistic amplification of masking and demasking effects. Furthermore, when this condensate structure enters the interaction region of high-energy hadron collisions, it can significantly alter the production rules of secondary pions, naturally yielding a gamma-ray spectrum with typical broken power-law characteristics. Consequently, GC provides a unified hadronic explanation for the observed BPL spectra from several ultra-high-energy astrophysical sources whilst also opening up new avenues for the reverse detection of the deep structure of protons using cosmic-ray signals. Although higher-order corrections, correlation function modelling, and hadronisation details still require further refinement, the core criteria of the GC solution and its few-parameter formulation ensure that this picture remains well-testable. In the future, with higher-precision cosmic gamma-ray observations, experiments at the LHC and next-generation hadron colliders, and advances in the study of pion saturation and condensation mechanisms, the theoretical foundations of the GC and its observational signatures are expected to undergo more rigorous and systematic testing.
    
\begin{acknowledgments}
We are grateful for the support from Hubei Polytechnic University. Y.-Y.Z. gratefully acknowledges the support from the Research Center for Ray Detection Discipline and Technology (Sichuan University of Science and Engineering).
\end{acknowledgments}

\section*{Abbreviations}
The following abbreviations are used in this manuscript:
\\

\noindent 
\begin{tabular}{@{}ll}
DGLAP & Dokshitzer--Gribov--Lipatov--Altarelli--Parisi\\
BFKL & Balitsky--Fadin--Kuraev--Lipatov\\
GLR-MQ & Gribov--Levin--Ryskin--Mueller--Qiu\\
BK & Balitsky-Kovchegov\\
JIMWLK & Jalilian-Marian--Iancu--McLerran--Weigert--Leonidov--Kovner\\
CGC & Color Glass Condensate\\
ZRS & Zhu--Ruan--Shen\\
ZSR & Zhu--Shen--Ruan\\
GC & Gluon Condensation\\
TOPT & Time-Ordered Perturbative Theory\\
QED & Quantum Electrodynamics\\
EBL & Extragalactic Background Light\\
PL & Power Law\\
BPL & Broken Power Law\\
LHAASO & Large Hadron Astrophysics Observatory\\
SED & Spectral Energy Distribution\\
VHE & Very High Energy\\
HAWC & High Altitude Water Cherenkov Observatory\\
CTA & Cherenkov Telescope Array\\
SWGO & Southern Wide-Field Gamma-ray Observatory\\
SNR & Supernova Remnant\\
AGN & Active Galactic Nuclei\\
GRB & Gamma-ray Burst
\end{tabular}

%\bibliographystyle{JHEP-mod.bst}
%\bibliography{refs}

\end{document}